\documentclass[a4paper,nobibnotes,nofootinbib]{revtex4}

\usepackage{amsmath,amssymb}
\usepackage{epsfig}
\usepackage{graphicx}
\usepackage{xspace}

\newcommand{\be}{\begin{equation}}
\newcommand{\ee}{\end{equation}}
\newcommand{\bL}{\begin{Large}}
\newcommand{\eL}{\end{Large}}
\newcommand{\ba}{\begin{eqnarray}}
\newcommand{\ea}{\end{eqnarray}}
\newcommand{\bc}{\begin{center}}
\newcommand{\ec}{\end{center}}
\newcommand{\bfig}{\begin{figure}}
\newcommand{\efig}{\end{figure}}
\newcommand{\f}[2]{\frac{#1}{#2}}
\newcommand{\g}{\gamma}
\newcommand{\om}{\omega}

\newcommand{\la}{\label}

\newcommand{\sg}{\sigma}

\newcommand{\al}{\alpha}

\newcommand{\rr}[4]{#1, {\it #2 \/}{\bf #3} #4}

\newcommand{\mW}{\ensuremath{m_{\mathrm{W}}}\xspace}
\newcommand{\mWfit}{\ensuremath{m_{\mathrm{W}}^{\mathrm{fit}}}\xspace}
\newcommand{\mWmin}{\ensuremath{m_{\mathrm{W}}^{\mathrm{min}}}\xspace}
\newcommand{\mt}{\ensuremath{m_{\mathrm{t}}}\xspace}
\newcommand{\ff}{\ensuremath{{\cal F}}\xspace}

\newcommand{\GeV}{\ensuremath{\mbox{GeV}}\xspace}

\newcommand{\Ifb}{\ensuremath{\mathrm{fb^{-1}}}\xspace}
\newcommand{\ttbar}{\ensuremath{\mathrm{t\bar{t}}}\xspace}
\newcommand{\WW}{\ensuremath{\mathrm{W^+W^-}}\xspace}
\newcommand{\W}{\ensuremath{\mathrm{W}}\xspace}
\newcommand{\ut}{\ensuremath{\mathrm{t}}\xspace}

\begin{document}
\title{Threshold scans in Central Diffraction at the LHC}

\author{M. Boonekamp}\email{boon@hep.saclay.cea.fr} 
\affiliation{Service de physique des particules, CEA/Saclay,
  91191 Gif-sur-Yvette cedex, France}
\author{J. Cammin}\email{cammin@fnal.gov} 
\affiliation{University of Rochester, New York, USA}
\author{R. Peschanski}\email{pesch@spht.saclay.cea.fr}
\affiliation{Service de physique th{\'e}orique, CEA/Saclay,
  91191 Gif-sur-Yvette cedex, France\footnote{%
URA 2306, unit{\'e} de recherche associ{\'e}e au CNRS.}}
\author{C. Royon}\email{royon@hep.saclay.cea.fr}
\affiliation{Service de physique des particules, CEA/Saclay,
  91191 Gif-sur-Yvette cedex, France, and Fermilab, Batavia, USA}

\begin{abstract}
We propose a new set of measurements which can be performed at
the LHC using roman pot detectors.
The method exploits excitation curves in central diffractive pair production, 
and is illustrated using the examples of the W boson and top quark mass measurements. 
Further applications are mentioned.
\end{abstract}

\maketitle

\section{Introduction}

We propose a new method to measure heavy particle properties via double 
photon and double pomeron exchange (DPE), at the LHC. In this category of events, the heavy objects 
are produced in pairs, whereas the beam particles
often leave the interaction region intact, and can be measured using very forward detectors.

If the events are $exclusive$, \emph{i.e.}, if no other particles are produced in addition to the pair of heavy objects 
and the outgoing protons, the proton measurement gives access to the photon-photon or pomeron-pomeron 
centre-of-mass \cite{Albrow:2000na}, and the dynamics of the hard process can be accurately studied. 
In particular, one can observe the threshold excitation and attempt to extract the heavy particle's mass, 
or study the particle's (possibly energy-dependent) couplings by measuring cross-sections and angular 
distributions. As examples of this approach, we give a detailed account of the \W boson and top quark 
mass measurement at production threshold. The method can easily be extended to other heavy objects. 

We also  consider {\it inclusive} double pomeron exchange, \emph{i.e.}, events where other 
particles accompany the heavy system. This production mode is to be 
expected in central diffractive \ttbar production, since inclusive 
double diffractive dijet production has effectively been observed at the 
Tevatron \cite{cdf}. The cross-section measurement of \cite{cdf}
allows for a 
rough estimate of the LHC expectation \cite{BPR}. The interest of such events is reviewed.

The paper is organised as follows. We start by giving the theoretical 
formulation  of \WW production (via QED) and of \ttbar production 
(in both exclusive and inclusive DPE). We then describe the event generation,
the simulation of detector effects, and the cuts used in the analysis.
The following part of the paper describes in detail the threshold scan method, 
in a twofold version (``turn-on''and ``histogram'' fits), and its application to 
the \W boson and top quark mass measurements. We then conclude on the method in
general, on the above mass measurements in particular, and mention a number of further applications.

\section{Theoretical formulation}

Pair production of \W bosons and top quarks in QED and  double pomeron exchange are described in detail in this section. 
\WW pairs are produced in photon-mediated processes, which are exactly calculable in QED. There is 
basically no uncertainty concerning the possibility of measuring these processes
at the LHC. On the contrary, \ttbar events, produced in 
exclusive double pomeron exchange, suffer from theoretical uncertainties since 
exclusive diffractive production is still to be observed at the Tevatron, 
and other models lead to different cross sections, and thus to a different
potential for the top quark mass measurement. However, since the exclusive 
kinematics are simple, the model dependence will be essentially reflected by 
a factor in the effective luminosity for such events. 

By contrast, the existence of inclusive double pomeron exchange --- in other 
words, when the pomeron remnants carry a part of the available center-of-mass 
energy --- is  certain since it has been observed already in  experiments. We will 
mention at the end how these events could be used and the interest of their 
experimental determination. We will briefly analyse their impact on
the \ttbar  threshold scan but we postpone a precise study of such events to a 
forthcoming publication.

\begin{figure} [ht]
\begin{center}
\epsfig{figure=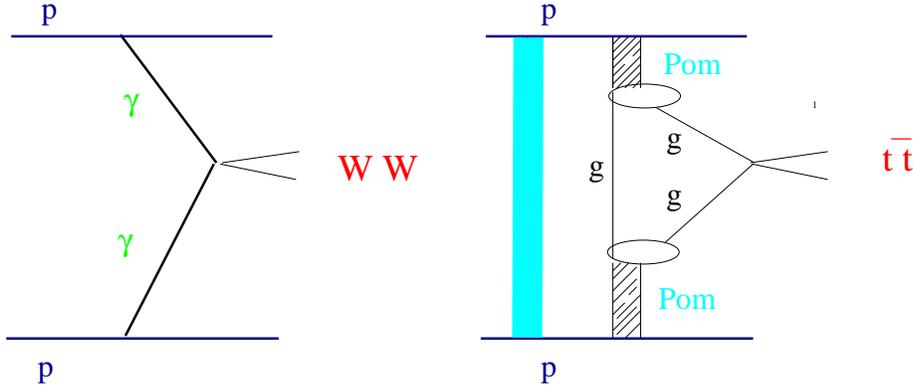,height=2 in}
\end{center}
\caption{{ \WW (QED) and \ttbar (QCD) exclusive production.} Left: 
double photon exchange process. Right: exclusive double pomeron exchange in the 
Bialas-Landshoff model via  ``gluons in the pomeron''; The grey band represents the 
rapidity gap survival suppression factor (see text).}
\label{fig1}
\end{figure}

\subsection{\boldmath \WW production via double photon exchange}

The QED process rates are obtained from  the following  cross section formula
\be
d\sg_{(\mathrm{pp\to \ p\ W^+ W^-p})} = \hat {\sg}_{\mathrm{\g\g\to W^+ W^-}}\ dn^{\g}_1 \ 
dn^{\g}_2\nonumber \ ,
\la{dsigmaWWW}
\ee
\
where the Born $\mathrm{\g\g\to W^+ W^-}$ cross-section reads 
\cite{Papageorgiu:1990mu}
\be
\hat {\sg}_{\mathrm{\g\g\to W^+ W^-}}=\frac {8\pi\al^2}{M_{\mathrm{WW}}^2}\left\{
\f 1t\left(1+\f34t+3t^2\right)\Lambda-3t(1-2t)
\ln\left(\f{1+\Lambda}{1-\Lambda}\right)\right\},
\la{hatsigma}
\ee
with
\be
t= \f{m_\W^2}{M_{\mathrm{WW}}^2}\ ,\ \ \ \ \ \ \ \ \ \ \Lambda =\sqrt{1-4t}\ , 
\la{kine}
\ee
where $M_{\mathrm{WW}}$ is the total \WW mass. The photon fluxes $dn^{\g}$ are given 
by  
\cite{Budnev:1975zs}
\be
 dn^{\g}=\f{\al}{\pi} \f{\om}{\om}\left(1-\f{\om}{E}\right)
\left[\phi\left(\f{q^2_{max}}{q^2_{0}}\right)-\phi\left(\f{q^2_{min}}{q^2_{0}
}
\right)\right]\ ,
\ee
where
\be
\phi\left(x\right)\equiv (1+ay)
\left[\ln\left(\f x{1+x}\right)+\Sigma_{k=1}^3 \f1{k(1+x)^k}\right]
-\ \f{(1-b)y}{4x(1+x)^3+c(1+\f14 y)}
\left[\ln\left(\f {2+2x-b}{1+x}\right)+\Sigma_{k=1}^3 
\f{b^k}{k(1+x)^k}\right]\ ,
\la{flux}
\ee
and
\be
q^2_{0} \sim 0.71\ \mathrm{GeV}^2\ \ ;\ y= \f{\om^2}{E(E-\om)}\ \ ;\ a \sim 7.16\ \ ;\ 
b\ \  
\sim -3.96\ \ ;\ c\sim 0.028 \ , 
\la{kineflux}
\ee
in the usual dipole approximation for the proton electromagnetic form 
factors. $\om$ 
is the photon energy in the laboratory frame, $q^2$ the modulus of its mass 
squared in 
the range
\be
\left[q^2_{min},\ q^2_{max}\right]\equiv \left[\f{m^2\om^2}{E(E-\om)}\ , 
\f{t_{max}}{q^2_{0}}\right]\ ,
\la{range}
\ee  
where $E$ and $m$ are  the energy and mass of the incident particle and  
$t_{max}\equiv ({m_\W^2}/{M_{\W\W}^2})_{max} $ is defined by the experimental 
conditions.

The QED cross section $d\sg (\mathrm{pp\to \ p\ W^+ W^-p})$ is  a theoretically 
clear 
prediction. One should take into account however, two sources of (probably 
mild) 
correction factors. One is due to the soft QCD initial state radiation 
between 
incident protons which could destroy the large rapidity gap of the QED 
process. It 
is present but much less pronounced than for the rapidity gap survival for a 
QCD 
hard process (see the  discussion in the next subsections), thanks to 
the large 
impact parameter implied by the QED scattering.  
 The 
second factor   is the QCD $\mathrm{gg \to  W^+ W^-}$ exclusive production via 
higher 
order diagrams. This has been evaluated recently  \cite{Binoth:2005ua} for 
standard 
(non diffractive) production to give a $5\%
$ correction factor. The similar
calculation for 
the diffractive \WW production by comparison with the QED process is outside 
the 
scope of our paper but deserves to be studied together with the ``inclusive'' 
background (\WW{}+hadrons) it could generate.

\subsection{\boldmath Exclusive diffractive production of \ttbar events}

Let us introduce the model \cite{bi90,ourpap} we shall use for describing 
exclusive \ttbar production in double diffractive production. 
This process is depicted in 
Fig.~\ref{fig1} (right). 

In \cite{bi90}, the diffractive mechanism is based on two-gluon exchange 
between 
the 
two incoming protons. The soft pomeron is seen as a  pair of gluons 
non-perturbatively
coupled  to the proton. One of the gluons is then coupled 
perturbatively to the hard process (either the Higgs boson, or the \ttbar 
pair, see Fig.~\ref{fig1}), while the other one plays the r\^ole of a soft 
screening of 
colour, 
allowing for diffraction to occur.
The corresponding cross-sections for $\mathrm{q \bar{q}}$ and Higgs boson production 
read:

\be
d\sigma_{\ttbar}^{exc}(s) = C_{\mathrm{q \bar{q}}}^{exc} 
\left(\frac{s}{M_{\mathrm{q \bar{q}}}^{2}}\right)^{2\epsilon}
                    \delta^{(2)}\left( \sum_{i=1,2} (v_{i} + k_{i}) \right)
                    \prod_{i=1,2}  d^{2}v_{i} d^{2}k_{i} d\eta_{i}\ 
                    \xi_{i}^{2\alpha'v_{i}^{2}}\! \exp(-2\lambda_{\ttbar} 
v_{i}^{2})
                    \ {\hat \sigma_{\ttbar}}\ ,
                    \label{exclusif}
\ee
where the variables $v_{i}$ and $k_{i}$ respectively denote the transverse
momenta of the outgoing protons and of top quarks, $\xi_{i}$ are the proton
fractional momentum losses,
and $\eta_{i}$ are the quark rapidities, 
\begin{equation}
\hat {\sigma}_{\ttbar} \equiv \frac{\pi}{24} \frac{d\sigma}{dt} = 
\frac{\rho(1-\rho)}{m_{T1}^{2}m_{T2}^{2}} \,\, , 
\,\,
\rho = \frac{4m_{\ut}^{2}}{M_{\ttbar}^{2}}\ 
\label{exc}
\end{equation} 
 is the hard $\mathrm{gg \to t \bar t}$  cross-section. 

In the model,  the soft pomeron 
trajectory is
taken 
from the standard Donnachie-Landshoff   parametrisation \cite{pom},
 namely $\alpha(t) = 1 + \epsilon + \alpha't$, with
$\epsilon \approx 0.08$ and $\alpha' \approx 0.25 \mathrm{GeV^{-2}}$. 
$\lambda_{\ttbar}$ and  the normalization  $C^{exc}_{\ttbar}$ are  
kept as in  the original paper \cite{bi90}. Note that, in this model, the 
strong (non perturbative) coupling constant is fixed to a reference value 
$G^2/4\pi =1,$ reflecting the lack of knowledge of the absolute normalization of 
exclusive DPE processes.

\subsection{\boldmath Inclusive diffractive production of \ttbar events}

It is convenient to introduce also the model for central inclusive  diffractive 
production \cite{BPR} applied to \ttbar dijets. One writes
\be
d\sigma_{\ttbar}^{incl} = C_{\ttbar}\left(\frac {x^g_1x^g_2 s }{M_{\ttbar}
^2}\right)^{2\epsilon}\!\! \! \! \delta ^{(2)}\! 
\left( \sum _{i=1,2}
v_i\!+\!k_i\right) \prod _{i=1,2} \!\!\left\{{d\xi_i}  {dx_i^g}
d^2v_i d^2k_i {\xi _i}^{2\alpha' v_i^2}\!
\exp \left(-2 v_i^2\lambda_{\ttbar}\right)\right\} 
\ {\sigma_{\ttbar}}\ G_P(x^g_1,\mu) G_P(x^g_2,\mu) \ .
\label{dinclujj}
\ee
\noindent In the above, $x_i^g$ are
the pomeron's momentum fractions carried by the gluons or quarks involved in 
the hard process, and the $G_P$ is the gluon 
energy density in the pomeron, $i.e.$, the gluon density multiplied by $x_i^g $. We use 
as 
parameterisations of the pomeron structure functions the fits to the 
diffractive 
HERA data 
performed in \cite{ba00}. The normalization  $C_{\ttbar}$ is 
obtained  from the description  \cite{BPR} of the jet-jet diffractive 
cross-section at the Tevatron \cite{cdf}.
The hard  cross-section ${\sigma_{\ttbar}}$ to be considered is now
\be 
{\sigma}_{\ttbar} = \frac{\rho}{{(m^{\ut}_{T1}})^{2}\ 
{(m^{\ut}_{T2}})^{2}}\left(1-\frac{\rho}{2}\right)\left(1-\frac{9\rho}{16}
\right)\ ,
\la{inclu_hard}
\ee 
to be distinguished from $\hat {\sigma}_{\ttbar}$  (\ref{exc}) due to inclusive 
characteristics \cite{BPR}.

\subsection{Rapidity Gap Survival}

In order to select exclusive diffractive states, such as for \WW (QED) and 
${\ttbar}$ (exclusive, QCD), it is required to take into account the 
corrections from soft hadronic scattering. Indeed, the soft scattering  
between incident particles tends to mask the genuine
hard diffractive interactions at 
hadronic colliders. The formulation of this correction \cite{sp} to the
scattering amplitude ${\cal A}_{(\mathrm{WW},\ttbar)}$ consists in considering a gap 
survival  probability ($SP$) function $S$ such that 
\begin{equation}
{\cal A}(p_{T1},p_{T2}, \Delta \Phi) =
\left\{ 1 +{\cal A}_{SP} \right\}{\bf \times} {\cal A}_{(\mathrm{WW},\ttbar)}\equiv {\cal S} 
{\bf \times} 
{\cal A}_{(\mathrm{WW},\ttbar)} = \int d^2{\bf k}_T\ {\cal S}({\bf k}_T) \ {\cal 
A}_{(\mathrm{WW},\ttbar)}({\bf p}_{T1}\!-\!{\bf k}_T,
{\bf p}_{T2}\!+
\!{\bf k}_T) 
\ ,  
\label{sp}
\end{equation}
where ${\bf p}_{T1,2}$ are the transverse momenta of the outgoing $p,\bar p$ 
and $\Delta \Phi$ their 
azimuthal angle separation.  
${\cal A}_{SP}$ is the soft 
scattering amplitude. 

The correction for the QED process 
is present but much less pronounced than for the rapidity gap survival for a 
QCD 
hard process, thanks to 
the large 
impact parameter implied by the QED scattering. In a specific model 
\cite{Khoze:2001xm} the correction factor  has 
been 
evaluated to be of order $0.9$ at the LHC for $\g\g\to \mathrm{H}.$
It is evaluated to be of order $0.03$ for the QCD exclusive diffractive 
processes at the LHC.

\section{Experimental context}

\subsection{The DPEMC Monte Carlo}
A recently developed Monte-Carlo program, {\tt DPEMC} \cite{dpemc}, provides 
an implementation of the \WW and \ttbar events described above in the
QED and both exclusive and inclusive double pomeron exchange modes.
It uses {\tt HERWIG} \cite{herwig} as a cross-section library of
hard QCD 
processes, and when required, convolutes them with the relevant pomeron 
fluxes and parton densities. The survival probabilities discussed in the 
previous section 
(respectively 0.9 for double photon  and 0.03 for double pomeron exchange 
processes)
have been introduced at the generator level. The cross section at the generator
level for \WW QED  and exclusive diffractive
\ttbar production is found to be 55.9 fb and 40.1 fb for a $m_\W$ mass of 
80.42
GeV and a top mass of 174.3 GeV after applying the survival probabilities.

\subsection{Roman pot detector positions and resolutions}

A possible experimental setup for forward proton detection is described in 
detail in
\cite{helsinki}. We will only describe its main features here and discuss its
relevance for the \W boson and top quark masses measurements.

In exclusive DPE or QED processes, 
the mass of the central heavy object can be reconstructed
using the roman pot detectors and tagging both protons
in the final state at the LHC. It is given  by $M^2 = \xi_1\xi_2 s$, where 
$\xi_i$ are 
the proton fractional momentum losses, and $s$  the total center-of-mass 
energy squared. In order to reconstruct objects with masses in the 160 GeV
range (for \WW events) in this 
way, the acceptance should be large down to $\xi$ values as low as a few 
$10^{-3}$. For \ttbar events, an acceptance down to $10^{-2}$ is
needed.
The missing mass resolution directly depends on the resolution on 
$\xi$, and should not exceed a few percent to obtain a good mass resolution.

These goals can be achieved if one assumes two detector stations, located at 
$\sim 
210$ m, and $\sim 420$ m \cite{helsinki} from the interaction point\footnote{A third 
position at 308 m is often considered as well but is more 
difficult from a technological point of view at the LHC and was not considered
for this study.}. The $\xi$ acceptance and 
resolution have been derived for each device using a complete simulation
of the LHC beam parameters. The combined $\xi$ acceptance is close to $\sim 60\%
$ at low masses (at about twice $m_\W$), and 90\% at higher masses
starting at about 220 GeV.   
for $\xi$ ranging from $0.002$ to $0.1$. The acceptance limit of the device 
closest to the 
interaction point
is $\xi > \xi_{min}=$0.02. Let us note also that the acceptance for \ttbar
events goes down to 20\% if only roman pots at 210 m are present since most
of the events are asymmetric (one tag at 420m and another one at 210m).

Our analysis does not assume any particular value for the $\xi$ resolution. 
We will discuss in the following how the resolution on the \W boson or
the top quark masses depend on the detector resolutions, or in other words,
the missing mass resolution.

\subsection{Experimental cuts}

This section summarises the cuts applied in the remaining part of the analysis.
As said before, both diffracted protons are required to be detected in 
roman pot detectors.

The triggers which will be used for the \WW and \ttbar events will be the
usual ones at the LHC requiring in addition a positive tagging in the roman
pot detectors.

The experimental offline cuts and their efficiencies have been obtained using a
fast simulation of the CMS detector \cite {cmsim} as an example, the fast
simulation of the ATLAS detector leading to the same results.
If we require at least one lepton (electron or muon) with a transverse
momentum greater than 20 GeV and one (two) jet with a transverse
momentum greater than 20 GeV (40 GeV) for \WW (\ttbar) to be
reconstructed in the acceptance of the main detector in addition to the tagged 
protons \footnote{The double pomeron exchange background to the signal is found
to be small, and will more correspond to misidentifications of jets as leptons
in the main detector. Since this is difficult to evaluate precisely using a fast
simulation of the detector, and this is quite small compared to the signal, we
decided not to incorporate it in the following study.},
we get an efficiency of about 50\% for \ttbar events, and 30\%
for \WW events. We give the mass resolution as a function of luminosity in the
following after taking into account these efficiencies. If the efficiencies are
found to be higher, the luminosities have to be rescaled by this amount.

\section{Threshold scan methods}

\subsection{Explanation of the methods}
We study two different methods to reconstruct the mass of heavy objects
double diffractively produced at the LHC. As we mentioned before, the method is
based on a fit to the turn-on point of the missing mass distribution at 
threshold. 

One proposed method (the ``histogram'' method) corresponds to the comparison of 
the mass distribution in data with some reference distributions following
 a Monte Carlo simulation of the detector with different input masses
corresponding to the data luminosity. As an example, we can produce 
a data sample for 100 fb$^{-1}$ with a top mass of 174 GeV, and a few 
MC samples corresponding to top masses between 150 and 200 GeV by steps of. 
For each Monte Carlo sample, a $\chi^2$ value corresponding to the 
population difference in each bin between data and MC is computed. The mass point 
where
the $\chi^2$ is minimum corresponds to the mass of the produced object in data.
This method has the advantage of being easy but requires a good
simulation of the detector.

The other proposed method (the ``turn-on fit'' method) is less sensitive to the MC 
simulation of the
detectors. As mentioned earlier, the threshold scan is directly sensitive to
the mass of the diffractively produced object (in the \WW case for instance, it
is sensitive to twice the \W mass). The idea is thus to fit the turn-on
point of the missing mass distribution which leads directly to the mass 
of the produced object, the \W boson. Due to its robustness,
this method is considered as the ``default" one in the following.

To illustrate the principle of these methods and their achievements,
we  apply them to the 
\W boson and the top quark mass measurements in the
following, and present in detail the reaches at the LHC. They can be applied to other 
threshold scans as well.

\subsection{W mass measurement}

\begin{figure}
  \centering
  \includegraphics[width=0.4\linewidth]{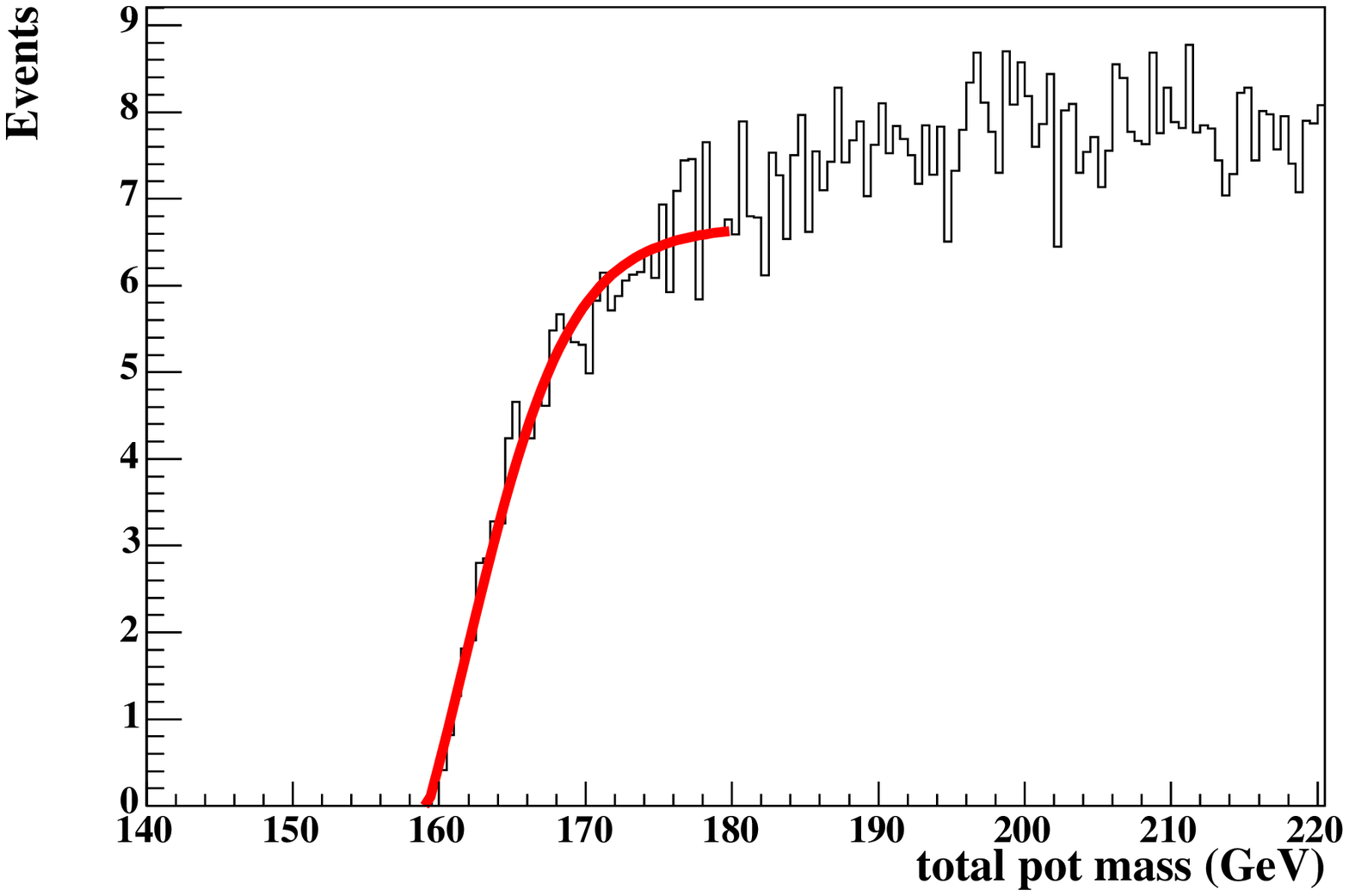}
  \includegraphics[width=0.4\linewidth]{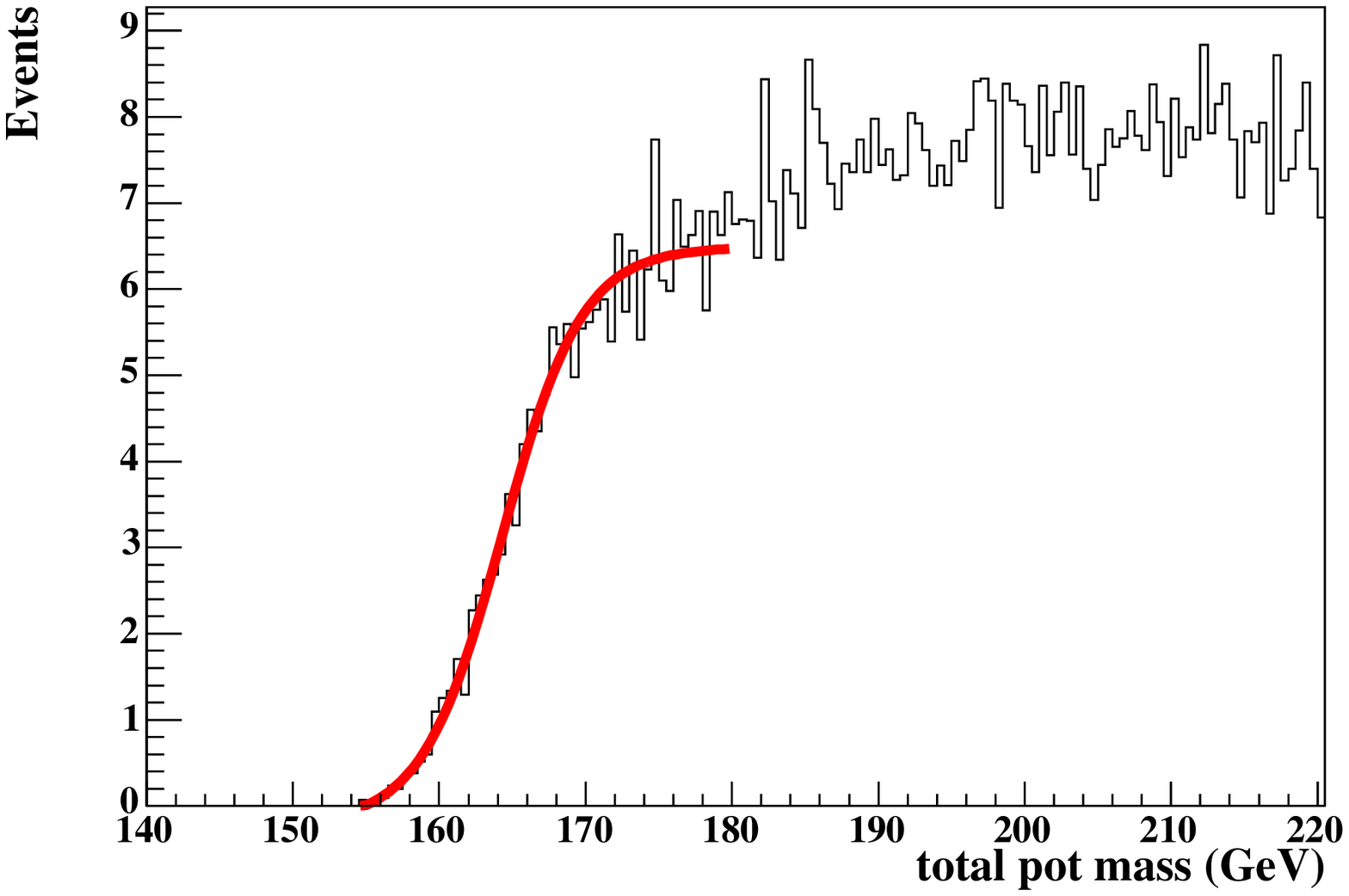}
  \caption{Two examples of fits to missing mass reference distributions with a
  resolution of the roman pot detectors of 1~\GeV (left) and 3~\GeV
  (right). We see on these plots the principle and the accuracy of the ``turn-on 
fits"
  to the MC at threshold.}
  \label{fig:WW_ref_fits}
\end{figure}

\begin{figure}
  \centering
  \includegraphics[width=0.4\linewidth]{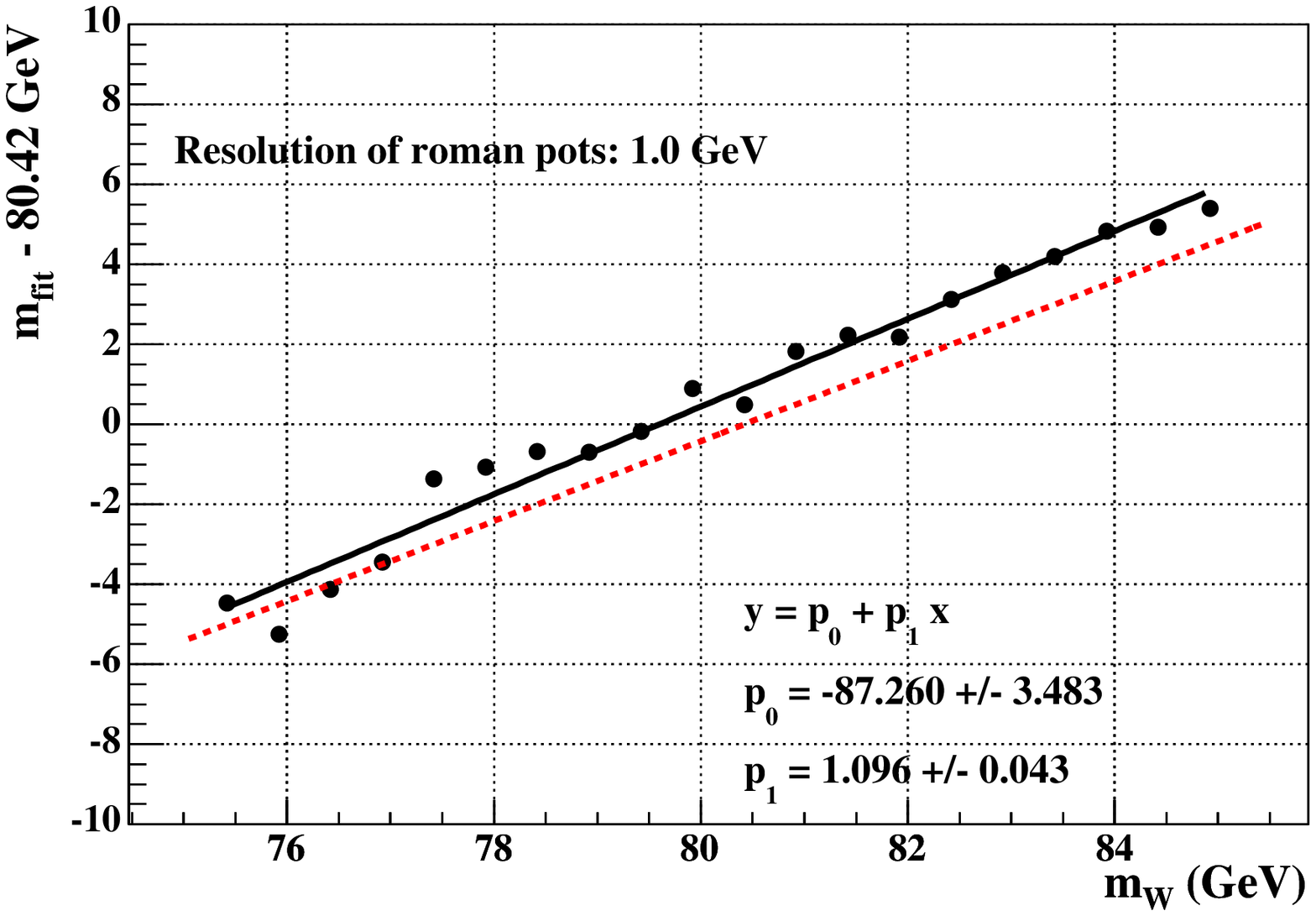}
  \includegraphics[width=0.4\linewidth]{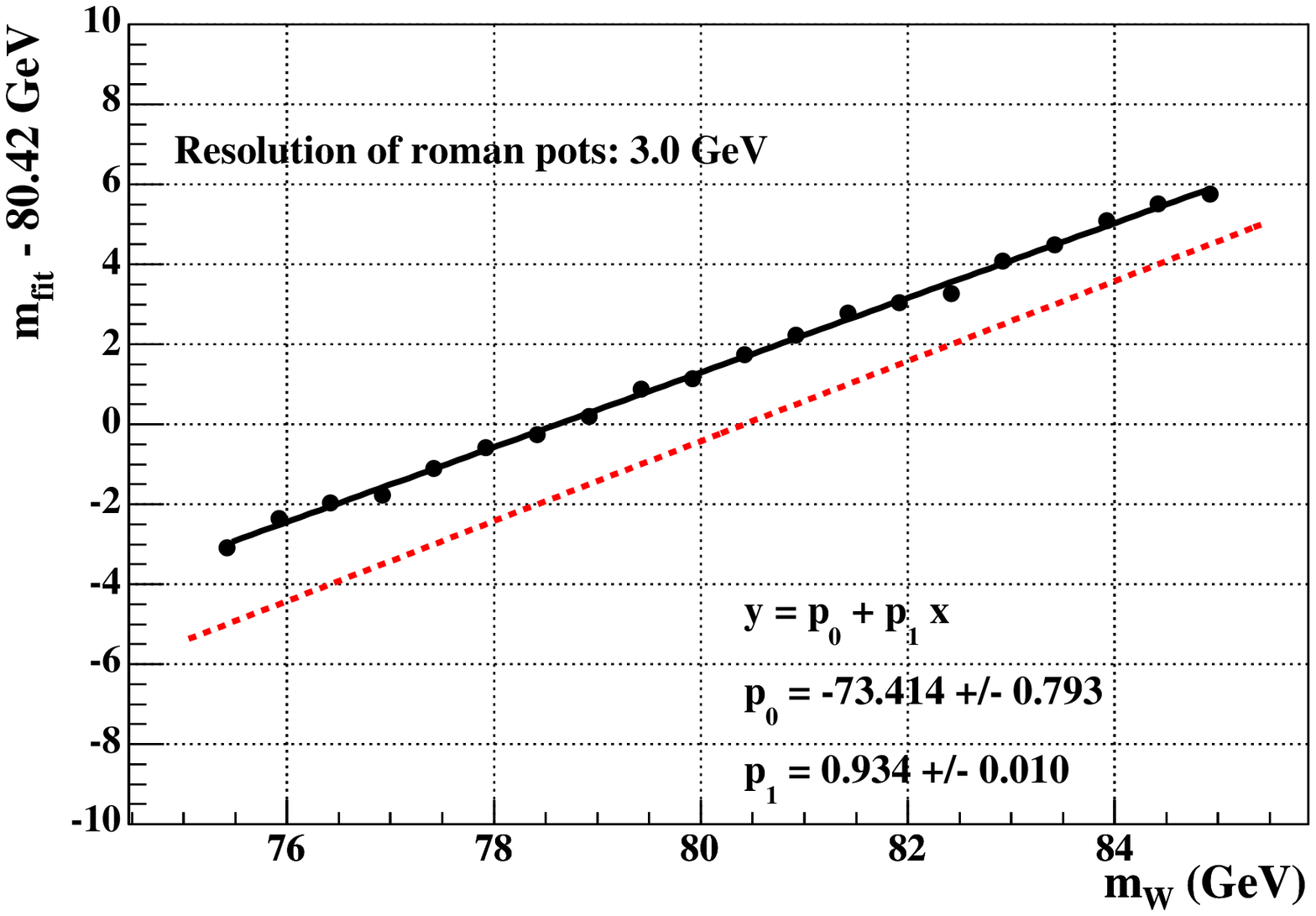}
  \caption{Calibration curves (see text) for two different roman pot resolutions 
of 1~\GeV
  (left) and 3~\GeV (right). We notice that the calibration can be fitted
  to a linear function with good accuracy. The dashed line indicates 
  the first diagonal to show the shift clearly.}
  \label{fig:calibration_WW}
\end{figure}

In this section, we will first describe the result of the ``turn-on fit" 
method to measure the \W mass. As we mentioned earlier, the advantage
of the \WW processes is that they do not suffer from any theoretical uncertainties
since this is a QED process.
The W mass can be extracted by fitting a 4-parameter `turn-on' curve to the 
threshold
of the mass distribution (c.f. Ref.~\cite{Abbiendi:2002ay}):
\begin{equation}\label{eq:fitfunc}
\ff =   P_1\cdot \left( \left[{e^{-\frac{x-P_2}{P_3}}+1}\right]^{-1}+P_4\right).
\end{equation}
$P_1$ is the amplitude, $P_2$ the inflexion point, $P_3$ the width of
the turn-on curve, and $P_4$ is a vertical offset, $x$ being
the missing mass. With a detector of
perfect resolution, $P_2$ would be equal to twice the \W mass.
However, the finite roman pot resolution leads to
a shift between $P_2$ and $2 m_\W$ which has to be established using a
MC simulation of the detector for different values of its resolution.
This shift is only related to the method itself and does not correspond
to any error in data. For each value of the \W input mass in MC, one has to 
obtain the
shift between the reconstructed mass ($P_2/2$) and the input mass, which
we call in the following the calibration curve.
It is assumed for simplicity that $P_2$ is a linear function of \mW,
which is a good approximation as we will see next. 
In order to determine the linear
dependence between $P_2$ and \mW, calibration curves are calculated
for several assumed resolutions of the roman pot detectors. The
calibration points are obtained by fitting \ff to the mass
distribution of high statistics samples (100\,000 events) for several
values of \mW. An example is given in Fig.~\ref{fig:WW_ref_fits} for
two resolutions of the roman pot detectors. The difference
between the fitted values of $P_2/2$ and the input \W masses
are plotted as a function of the input W mass and are then fitted with a
linear function.  To minimise the errors on the slope and offset, the difference 
$P_2/2-80.42~\GeV$ is plotted versus $\mW$
(Fig.~\ref{fig:calibration_WW}).

To evaluate the statistical uncertainty due to the method itself,
we perform the fits with some 100 different ``data" ensembles.
For each ensemble, one obtains a different 
reconstructed \W mass, the dispersion corresponding only to statistical
effects.
The expected statistical uncertainty on the actual measurement of the
W mass in data is thus estimated with these ensemble tests for several 
integrated
luminosities and roman pot resolutions. Each ensemble contains a
number of events that corresponds to the expected event yield for a
given integrated luminosity, taking into account selection and
acceptance efficiencies. The turn-on function \ff is fitted to each
ensemble. Only the parameters $P_1$ and $P_2$ are allowed to float,
$P_3$ and $P_4$ are fixed to the average values obtained from the fits
for the calibration points.

\begin{figure}
  \centering
  \includegraphics[width=0.35\linewidth]{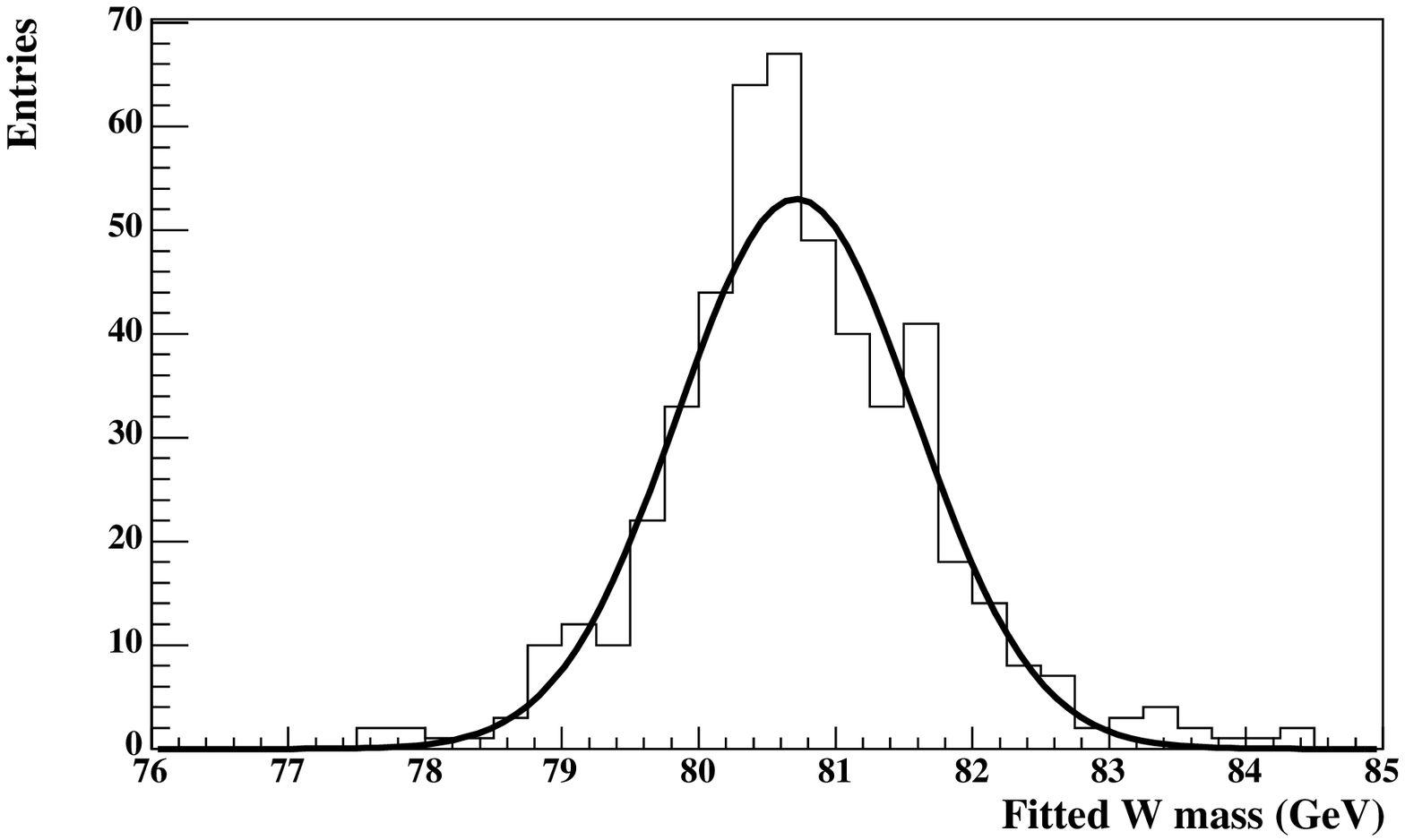}
  \qquad
  \includegraphics[width=0.35\linewidth]{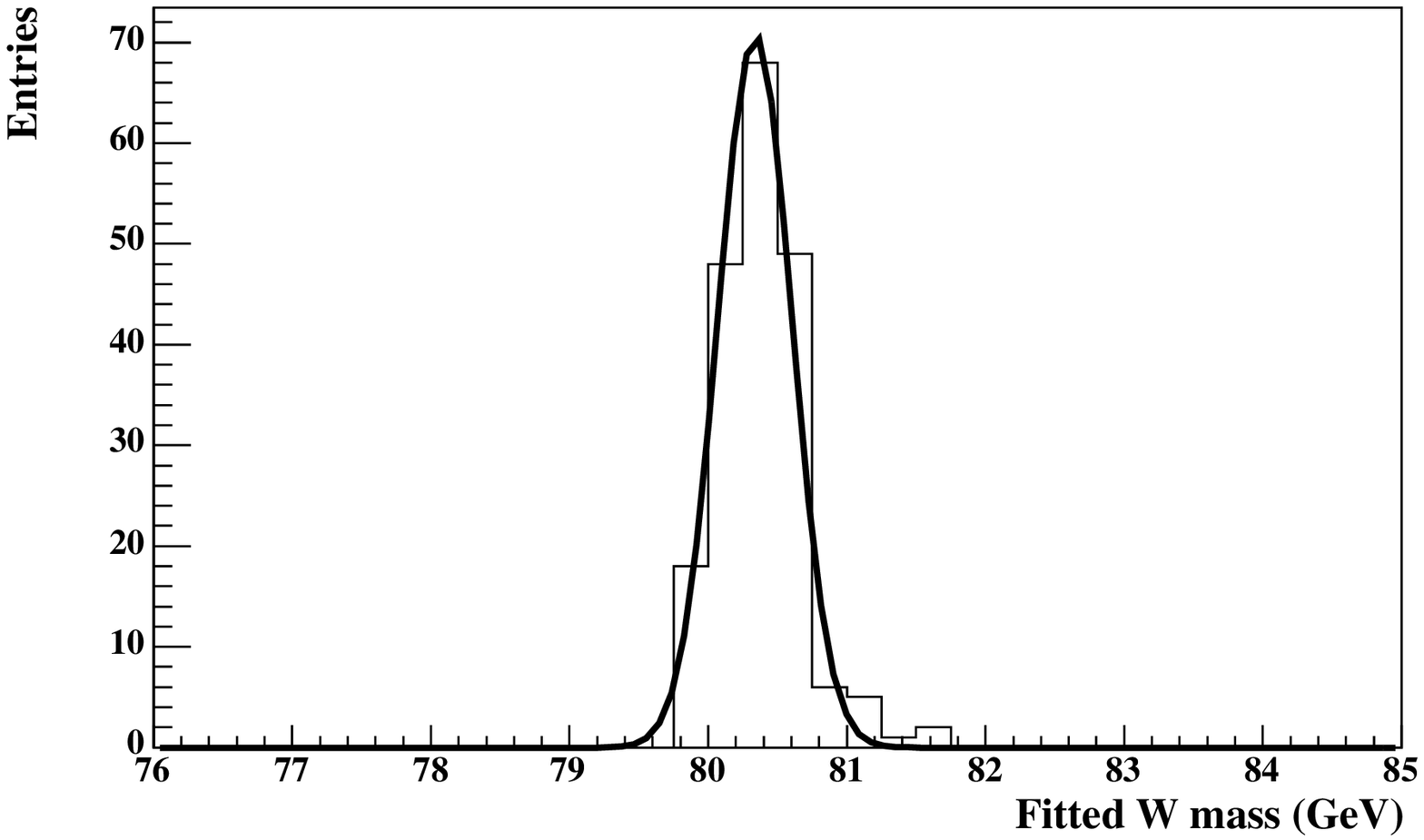}
  \caption{Distribution of the fitted value of the W mass from
  ensemble tests. Left: corresponding to 150~\Ifb , right:
  corresponding to 300~\Ifb.
  We note the resolution obtained on the \W mass for these two luminosities.}
  \label{fig:WW_ensemble_distributions}
\end{figure}

\begin{figure}
  \centering
  \includegraphics[width=0.35\linewidth]{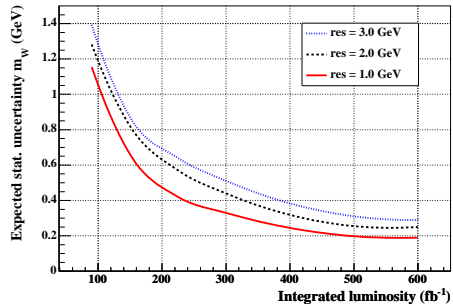}
  \caption{Expected statistical uncertainty on the \W mass
  as a function of luminosity for three different roman
  pot resolutions.
  }
  \label{fig:ww_resvslumib}
\end{figure}

In order to obtain the fitted estimate for the W mass, \mWfit, in each
ensemble, the fit value of $P_2$ is corrected with the calibration
curve that corresponds to the roman pot resolution. For each
resolution \mWfit is histogrammed as shown in
Fig.~\ref{fig:WW_ensemble_distributions}.  The distributions are
fitted with a Gauss function where the width corresponds to the
expected statistical uncertainty of the W mass measurement.
Fig.~\ref{fig:ww_resvslumib} shows the expected precision as a function
of the integrated luminosity for several roman pot resolutions.
With 150~\Ifb the expected statistical uncertainty on \mW is about
0.65~\GeV when a resolution of the roman pot detectors of 1~\GeV can be
reached. With 300~\Ifb the expected uncertainty on \mW decreases to about 0.3~\GeV.

We notice of course that this method is not competitive to get a precise
measurement of the \W mass, which would require a resolution to be better
than 30 MeV. However, this method can be used to align precisely the roman
pot detectors for further measurements. A precision of 1 GeV (0.3 GeV)
on the \W mass leads directly to a relative resolution of 1.2\%
(0.4\%) on $\xi$ using the missing mass method. This calibration
will be needed, for instance, to measure the top mass as proposed in the
next section.

\begin{figure}
  \includegraphics[width=0.3\linewidth]{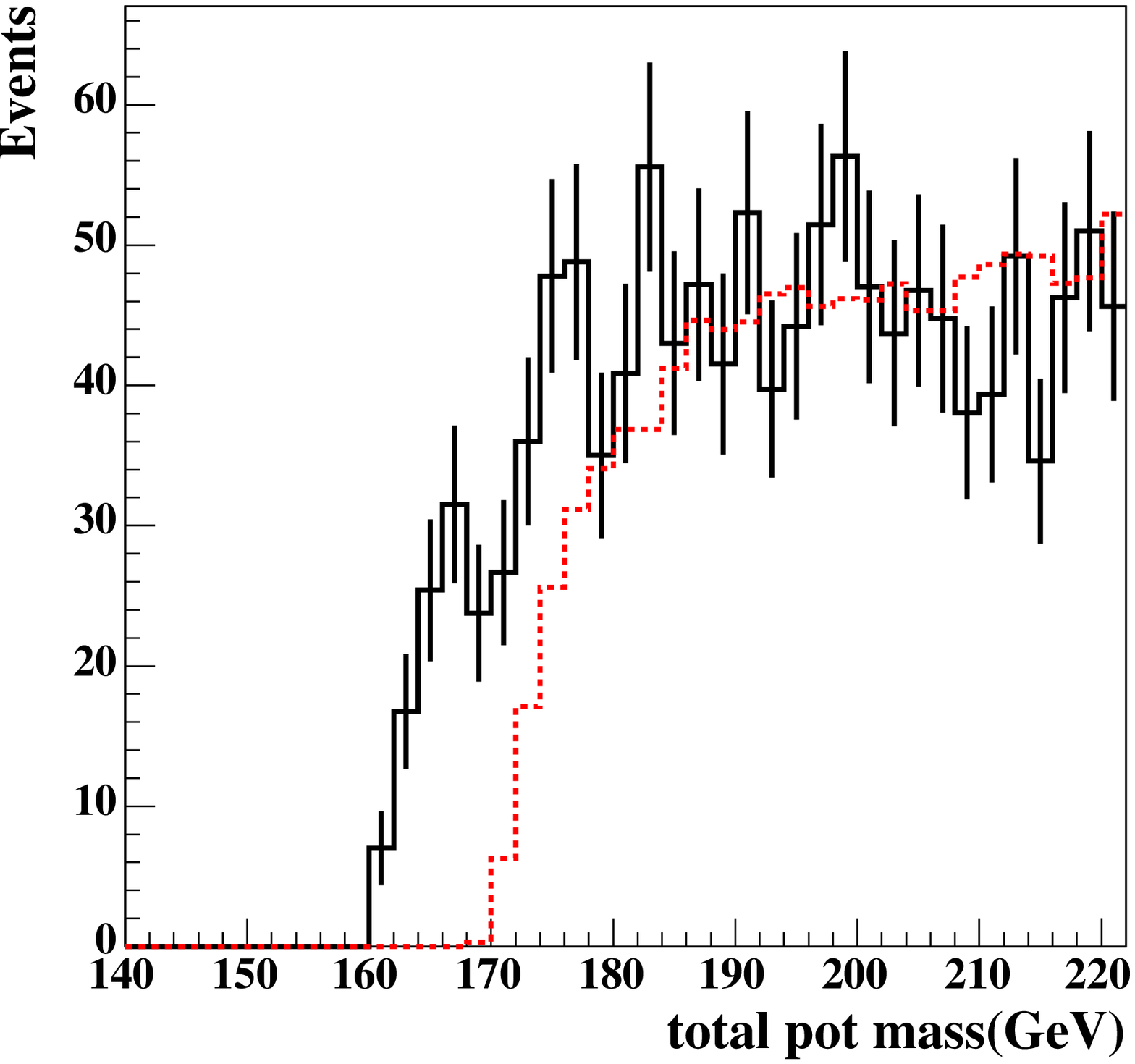}
  \qquad
  \includegraphics[width=0.3\linewidth]{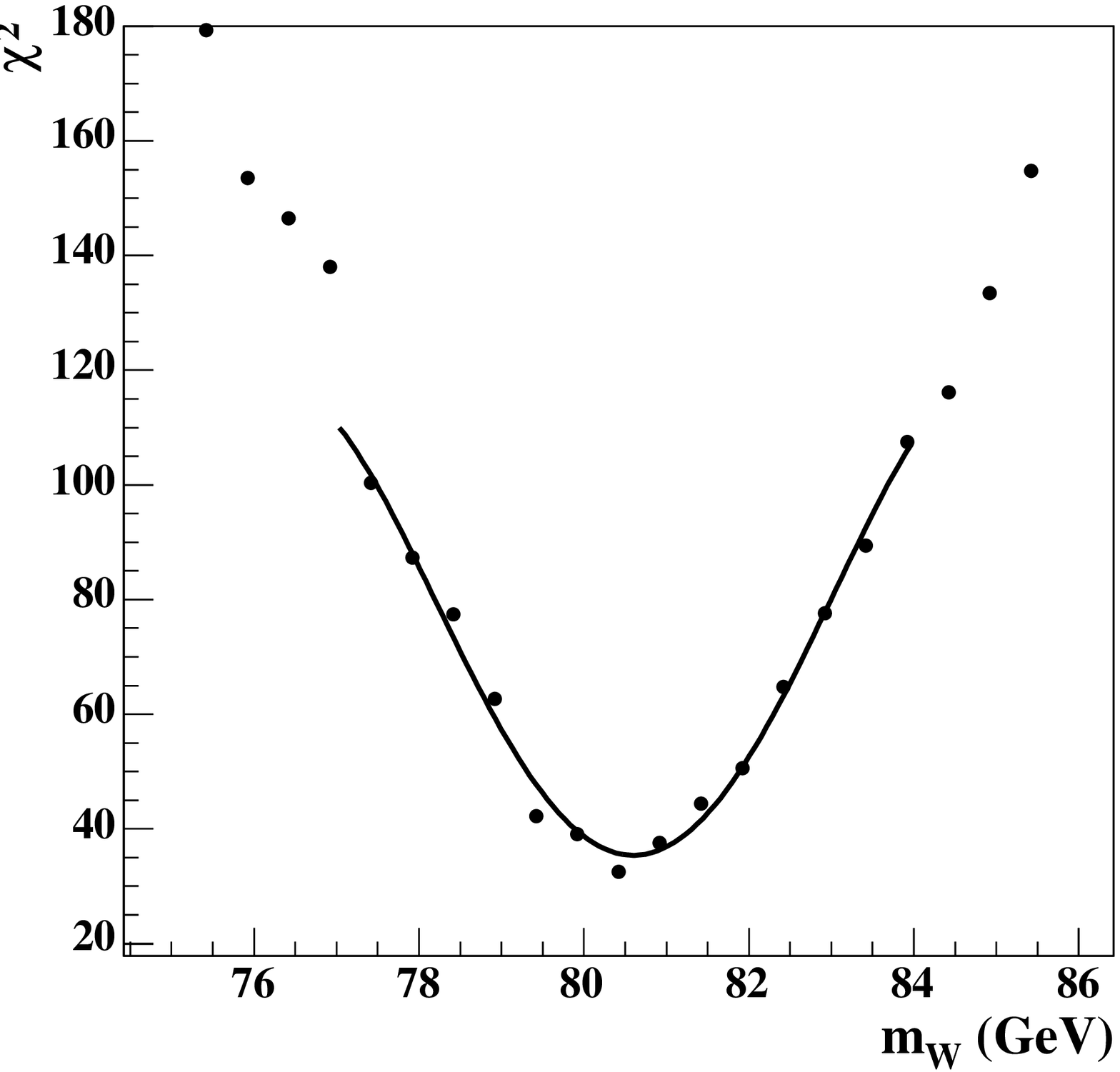}
  \caption{Left: Example of the histogram-fitting method. We see the difference
  between the ``data" sample (full histogram with error bars,
  $\mW=80.42~\GeV$) and a reference histogram (dashed line, $\mW=85.42~\GeV$). 
  Right:
    Example of the $\chi^2$ distribution in one ensemble.}
  \label{fig:W_mass_hist_fit}
\end{figure}

Let us now present the result on the ``histogram" method, which is an
alternative approach to determine the W mass.
The same high statistics templates used to derive the calibration
curves are fitted directly to each ensemble
(see Fig.~\ref{fig:W_mass_hist_fit} left). 
The $\chi^2$ is defined using the approximation of poissonian errors
as given in Ref.~\cite{Gehrels:1986mj}.  Each ensemble thus gives a $\chi^2$ 
curve
which in the region of the minimum is fitted with a fourth-order
polynomial (Fig.~\ref{fig:W_mass_hist_fit} right). The position of the
minimum of the polynomial, \mWmin, gives the best value of the W mass
and the uncertainty $\sigma(\mW)$ is obtained from the values where
$\chi^2 = \chi^2_\text{min} + 1$. The mean value of $\sigma(\mW)$
for all ensembles are quoted as expected statistical uncertainties.

\begin{figure}
  \includegraphics[width=0.35\linewidth]{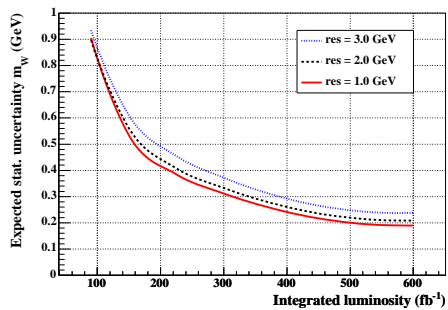}
  \caption{
    Expected statistical precision of the W mass as a function of the
    integrated luminosity for various resolutions of the roman pot
    detectors using the histogram-fitting method.}
  \label{fig:W_mass_hist_fitb}
\end{figure}

The expected statistical errors on the W mass using histogram fitting
are comparable to those using the function fitting method. However,
the former turns out to be more sensitive to the resolution of the roman pot
detectors.

\subsection{Top mass measurement}
\begin{figure}[tbhp]
  \centering
  \includegraphics[width=0.4\linewidth]{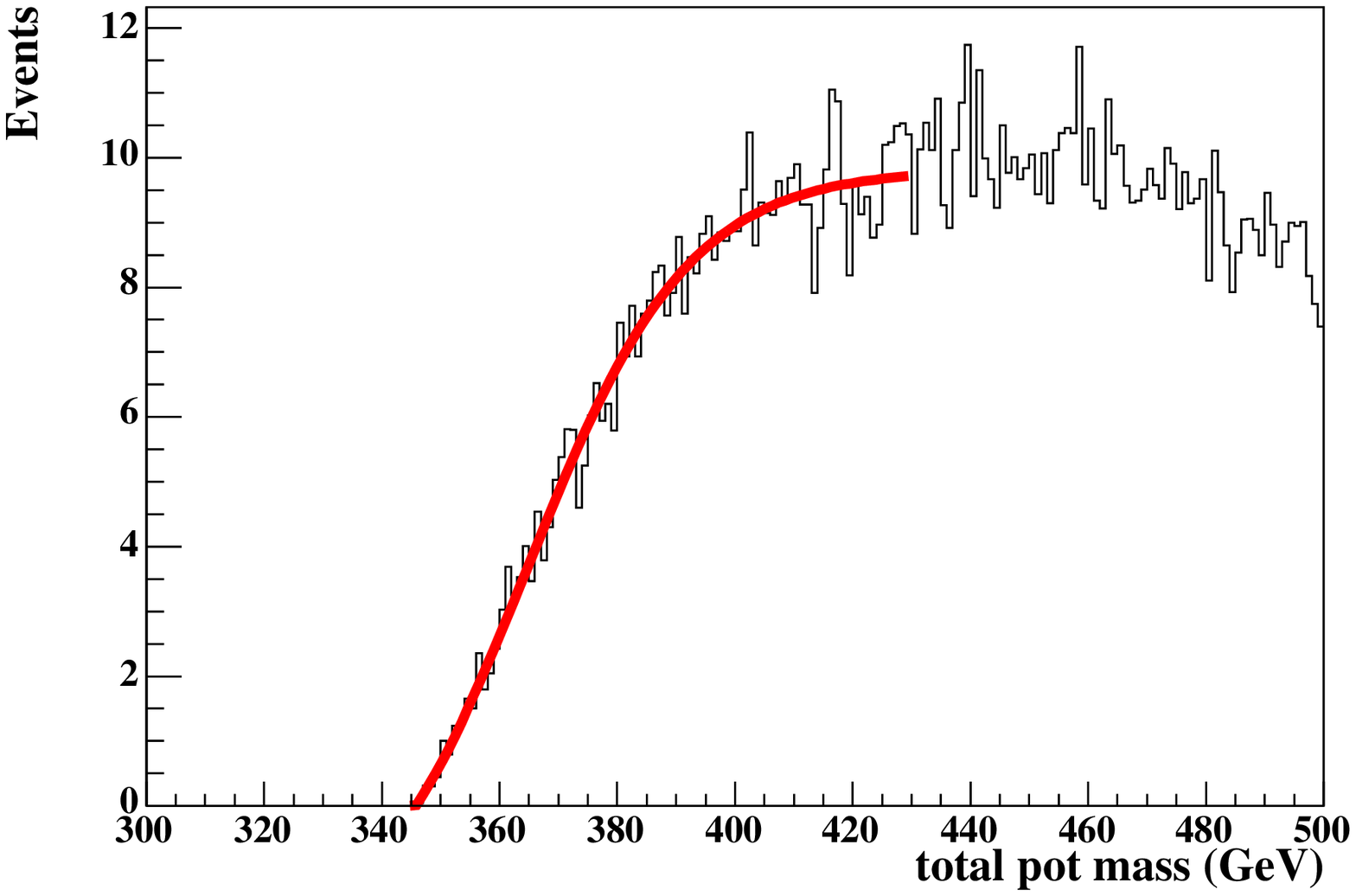}
  \qquad
  \includegraphics[width=0.4\linewidth]{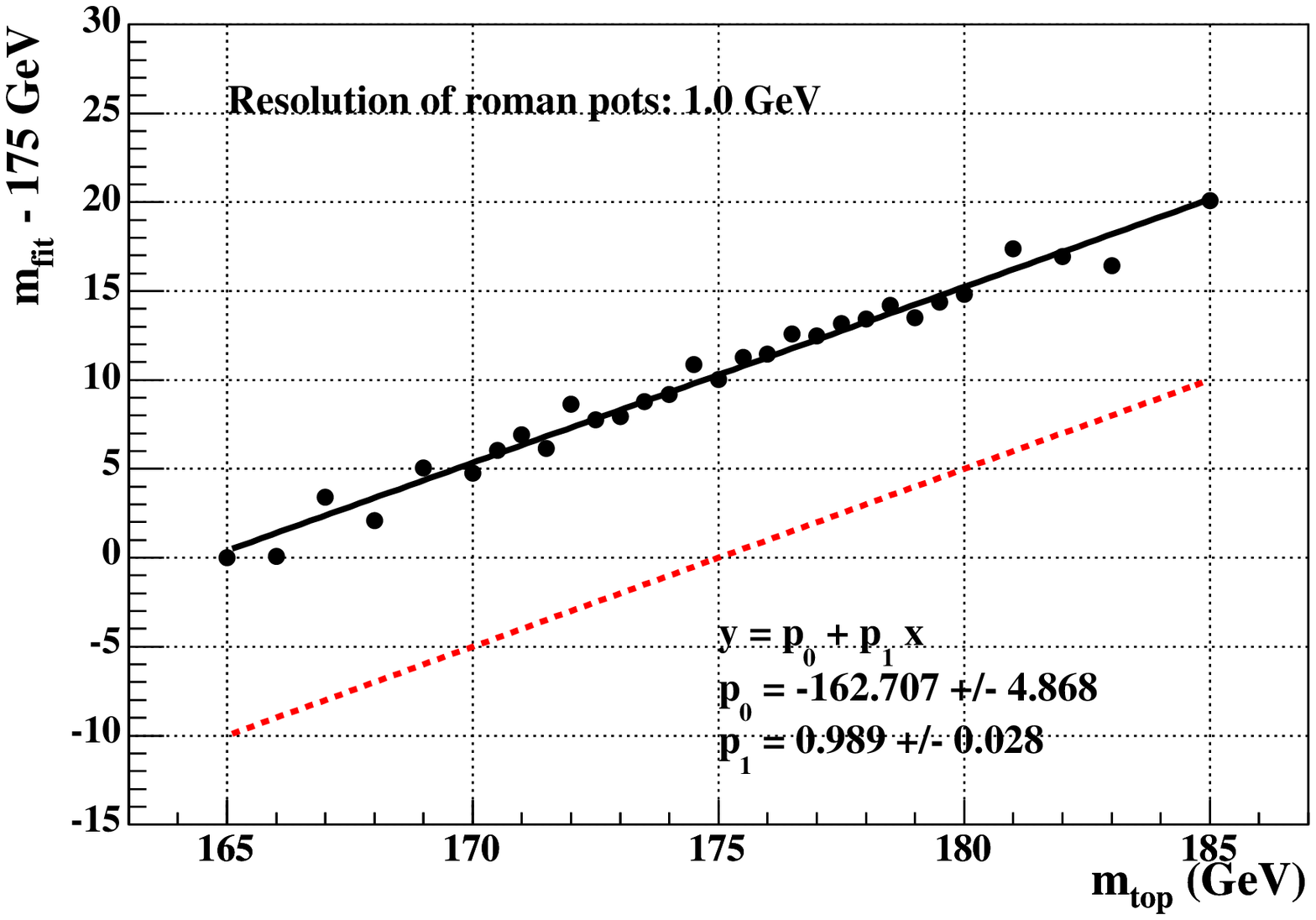}
  \caption{Fit to a reference mass distribution with $\mt=175~\GeV$
    (left) and calibration curve for a roman pot resolution of 1~\GeV,
    the diagonal is displayed in dashed line to show the difference (right).
    Note the larger difference between the calibration curve and the diagonal
    compared to the case of  \WW production.}
  \label{fig:calibration_ttbar}
\end{figure}

The method to extract the top mass is the same as for the W mass
described in the previous section. The theoretical cross section
is not as well known as for the \W and is model dependent. Our study
assumes the Bialas Landshoff model for exclusive \ttbar production.
For \ttbar events the width of the
turn-on curve is considerably larger than for WW events 
(Fig.~\ref{fig:calibration_ttbar}, left), resulting in a larger
offset between the actual turn on and the inflexion point of the fit
function\footnote{Note in addition that the top quark width
is not included in Herwig and thus in our study. However, this effect 
is expected to be small.}. The calibration curve for a resolution of the roman pot
detectors of 1~\GeV is displayed in Fig.~\ref{fig:calibration_ttbar},
right.

\begin{figure}
  \centering
  \includegraphics[width=0.4\linewidth]{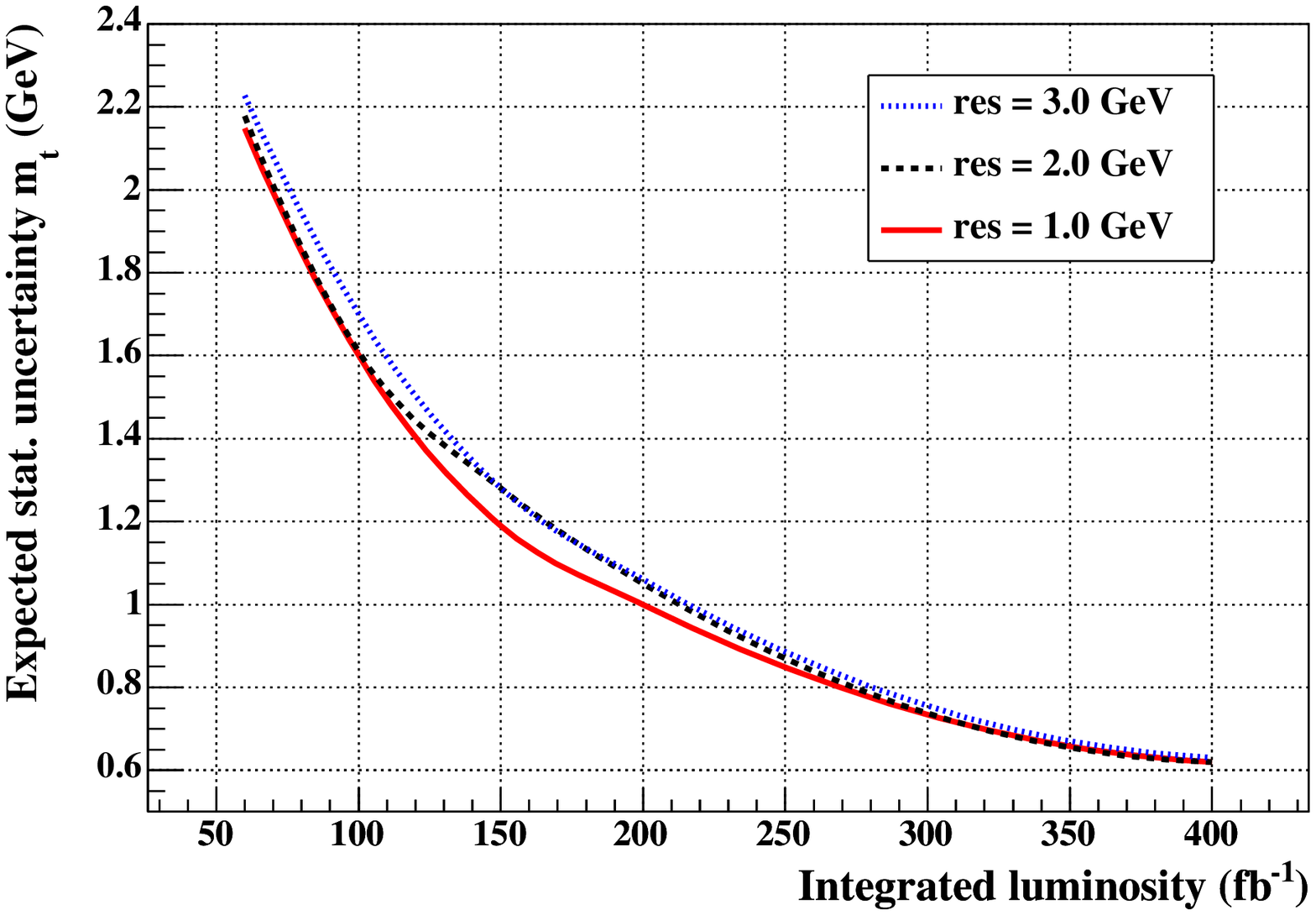}
  \includegraphics[width=0.4\linewidth]{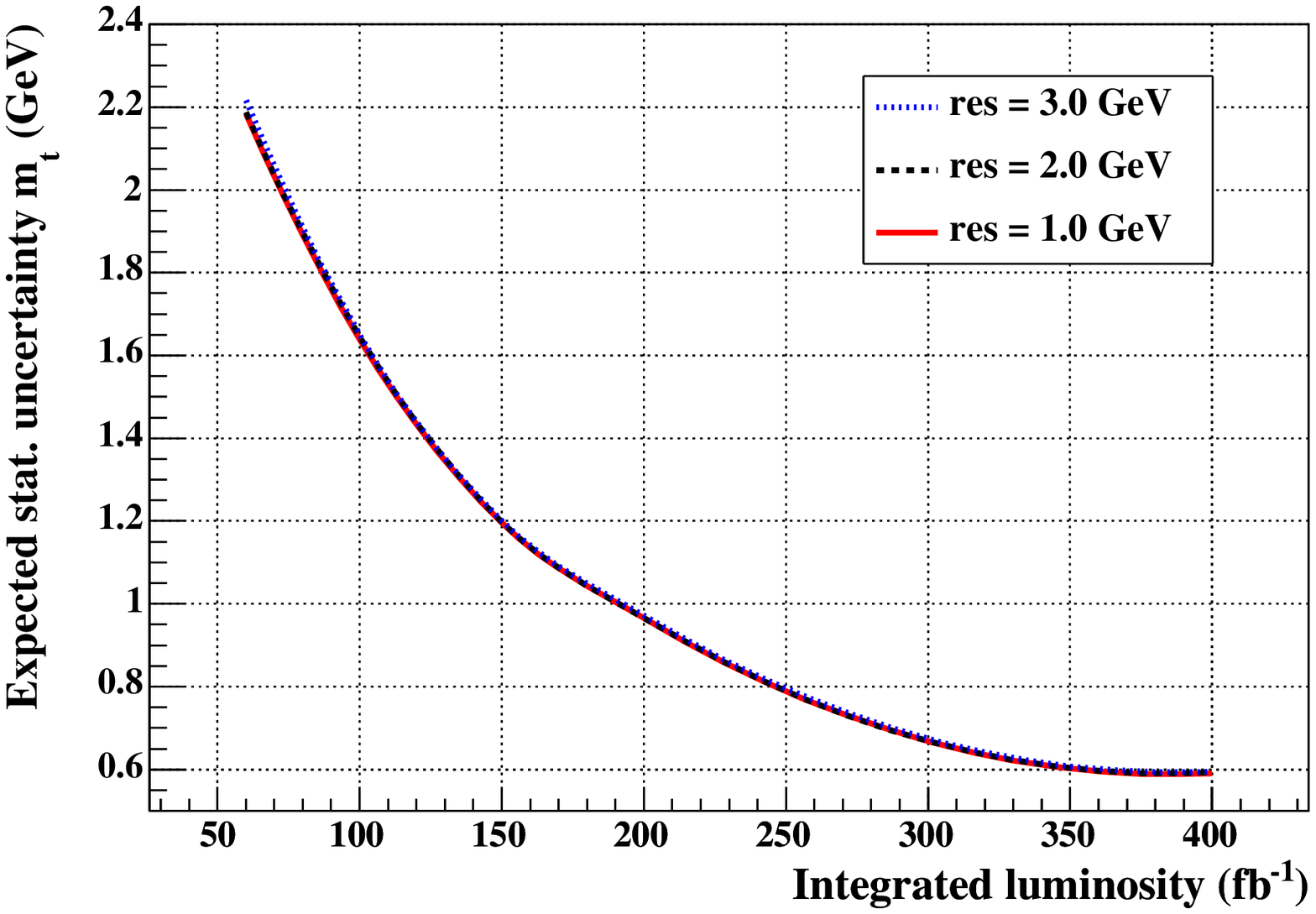}
  \caption{Expected statistical precision of the top mass
    as a function of the integrated luminosity for various resolutions
    of the roman pot detectors (full line: resolution of 1 GeV, dashed line: 2
    GeV, dotted line: 3 GeV). Left: function fitting method, right:
    histogram fitting method (the three curves for different roman pot
    resolutions lead to the same results and cannot be distinguished in the
    figure).}
  \label{fig:ttbar_resvslumi}
\end{figure}

Ensemble tests for integrated luminosities of 50, 75, 100 and
200~fb$^{-1}$ and roman pot detector resolutions of 1~\GeV, 2~\GeV and
3~GeV yield the results shown in
Fig.~\ref{fig:ttbar_resvslumi}, left. Resolutions of the roman pot detectors
between 1~\GeV and 3~\GeV give similar statistical uncertainties on
the top quark mass
which is due to the fact that the main limiting effect on resolution is
statistics. With 100~\Ifb the expected statistical precision is
about 1.6~\GeV and gets improved to about 0.65~\GeV with 300~\Ifb. 

The results have also been cross-checked using the histogram fitting
method which was found to yield very similar expected uncertainties as
the function fitting method (Fig.~\ref{fig:ttbar_resvslumi}, right).

\section{Outlook and prospects}

In this section, we discuss other applications of the threshold scan method.
Detailed analysis is postponed to forthcoming papers \cite{us}.

As we mentioned before, the cross section of exclusive top pair production at the LHC 
is still uncertain, and predictions will be constrained by the incoming results form the Tevatron, 
especially from the D\O\ experiment where it is possible to detect double tagged events. 
On the contrary, inclusive double pomeron exchange has
already been observed, and top quark pair production in this mode is fairly certain at the LHC.
In this case, the threshold excitation is sensitive to
quark and gluon densities at high pomeron momentum fraction, so that these events provide a rather 
unique opportunity to study structure functions near the endpoint.

\begin{figure} [ht]
\begin{center}
\epsfig{figure=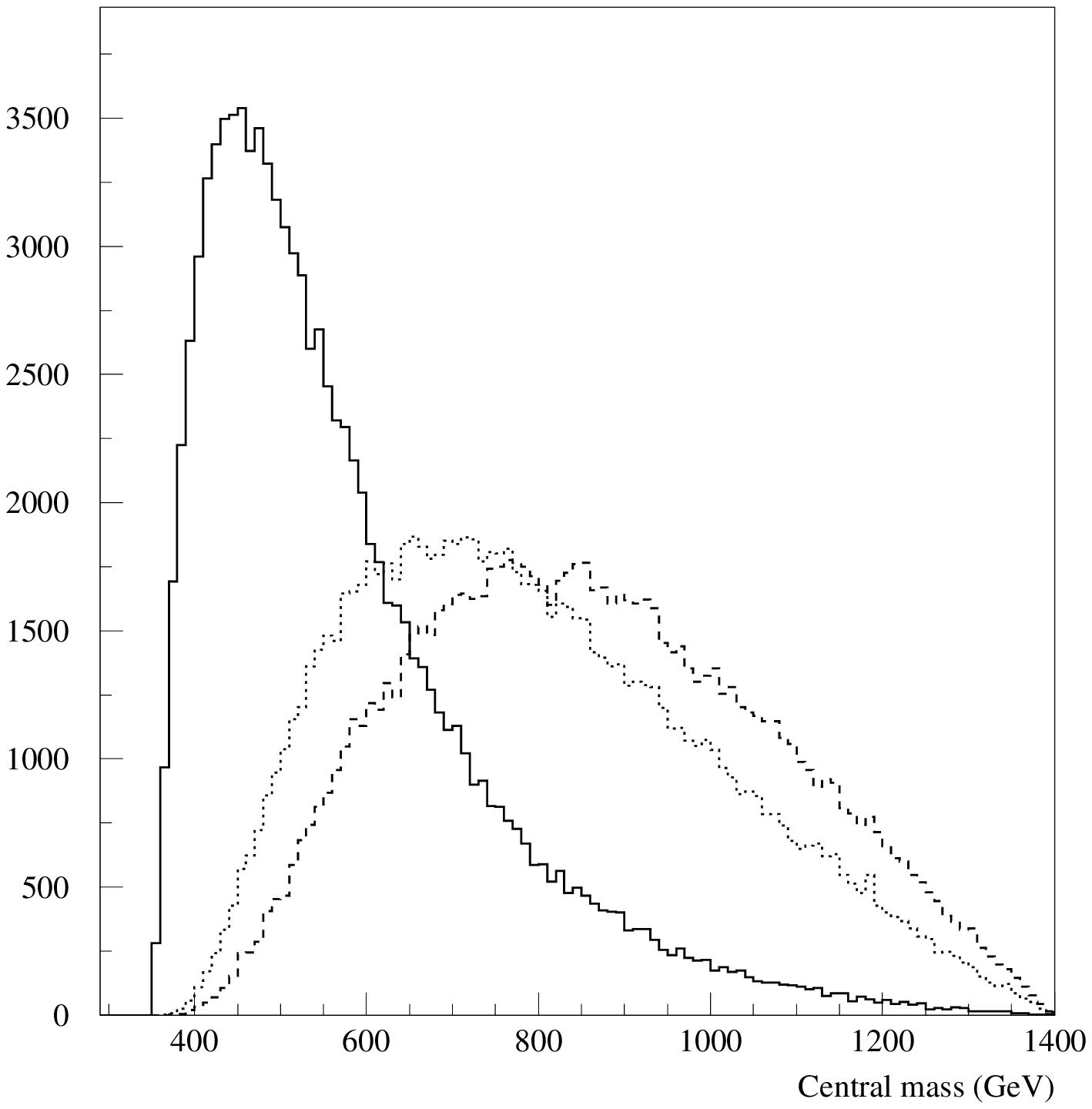,height=2.9 in}
\end{center}
\caption{{Missing mass
distributions at the generator level using the DPEMC Monte Carlo for
the exclusive \ttbar events in full line and the results on 
 inclusive \ttbar production for two different gluon
densities in the pomeron (dashed line: fit 1, dotted line: fit 2, see Ref.
\cite{h1}). } }
\label{fig1b}
\end{figure}

\begin{figure} [ht]
\begin{center}
\epsfig{figure=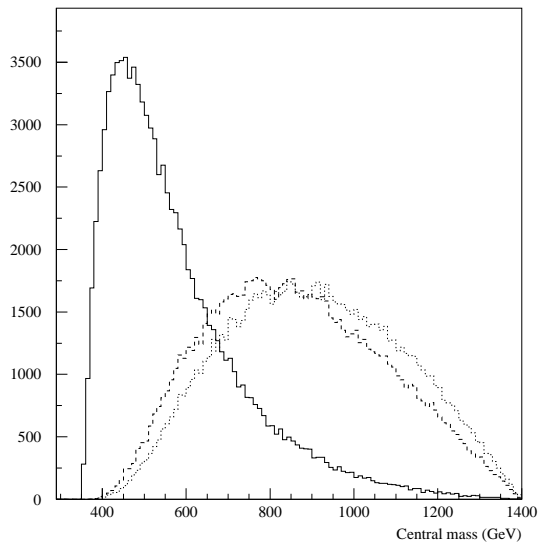,height=2.9 in}
\end{center}
\caption{{Missing mass
distributions at the generator level using the DPEMC Monte Carlo for
the exclusive \ttbar events in full line,
the inclusive events from model~\cite{BPR,ourpap} (dashed line), and
\cite{POMWIG} (dotted line), see text.}}
\label{fig2b}
\end{figure}

To illustrate this point, we give in Fig. \ref{fig1b} the missing mass
distributions at the generator level using the DPEMC Monte Carlo for
the exclusive \ttbar events (full line) and the results on the 
Bialas-Landshoff inclusive \ttbar production for two different gluon
densities in the pomeron (dashed line: fit 1, dotted line: fit 2, see Ref.
\cite{h1}). Fit 2 in Ref \cite{h1} leads to a more prominent gluon at high
$\beta$ than fit 1. We see that the missing mass distribution is directly
sensitive to the parton distributions in the pomeron. In Fig. \ref{fig2b},
we display the differences between the exclusive \ttbar events in full line
and the result of the factorisable POMWIG model (dotted line), and the non
factorisable one based on the Bialas-Landshoff approach. We see again that the
missing mass distribution, and thus the threshold analysis can help
distinguishing between the models.

Another application of exclusive pair-production consists in measuring the mass of stops 
and sbottoms, provided these particles exist and can be produced in pairs at the LHC.

Finally, \W pair-production in central diffraction gives access to the couplings 
of gauge bosons. Namely, as we mentioned already, \WW production in 
two-photon exchange is robustly predicted within the Standard Model. Any anomalous coupling
between the photon and the \W will reveal itself in a modification of the
production cross section, or by different angular distributions. Since the cross-section of this process is proportional
to the fourth power of photon-\W coupling, good sensitivity is expected.

\section{Conclusion}

Recent work on DPE has essentially focused on the Higgs boson search in the exclusive channel.
In view of the difficulties and uncertainties affecting this search \cite{ourpap}, we highlight new aspects of
double diffraction which complement the diffractive program at the LHC.

In particular, QED W pair production provides a certain source of
interesting diffractive events. Inclusive \ttbar production via double pomeron exchange is 
also an open channel and will provide interesting information on a poorly known aspect of 
structure functions. These robust channels will help and accompany the understanding of the more 
intriguing and challenging problem of exclusive double diffraction. 

In this paper, we have advocated the interest of threshold scans in double
pomeron exchange. This method considerably extends the physics program at the LHC.
To illustrate its possibilities, we described in detail the \W boson and the top quark
mass measurements.
The precision of the \W mass measurement is not competitive with other methods, but provides a very precise calibration 
of the roman pot detectors. The precision of
the top mass measurement is however competitive, with an expected precision better than 1 GeV at high luminosity.
Other promising applications remain to be investigated.

\end{document}